# ADMIRE vs. SAFIRE: Objective comparison of CT reconstruction algorithms and their noise properties

**Author Block:** I. Dalehaug[1], K. N. Bolstad[1], D. Aadnevik[1], S. Flataboe[1], H. E. S. Pettersen[1,2]

[1]Department of Oncology and Medical Physics, Haukeland University Hospital, Bergen, Norway
[2]Department of Physics and Technology, University of Bergen, Norway

## Abstract

**Purpose:** Siemens has developed several iterative reconstruction (IR) algorithms on their CT scanners. SAFIRE is available on most of their CT scanners. The latest algorithm, ADMIRE, is available on their newest high-end CT scanners. The aim of our study was to compare the two IR algorithms' noise reduction properties using objective methods.

**Methods and Materials:** The homogeneous module of the Catphan phantom was scanned on a Siemens AS+ and a Siemens Flash CT scanner using an axial abdomen protocol with fixed tube current at two dose levels. The images were reconstructed with an abdomen filter (B30) using filtered back projection (FBP) and a low, medium, and high level of SAFIRE or ADMIRE. Noise Power Spectrum (NPS) curves were calculated using these images. Then, an anthropomorphic abdomen phantom (Kyoto Kagaku PH-5) was scanned using the same setup and exposure parameters. Fifty axial images at the same slice location were used to calculate inter-image standard deviation maps.

**Results:** At full dose, the median values of the NPS curves were similar for both scanners at all IR levels. At low dose the median values of the NPS curves were generally shifted towards lower spatial frequencies, usually resulting in a more "blotchy" image texture. This shift was more prominent for ADMIRE compared to SAFIRE for all IR levels. Based on the inter-image standard deviation maps of the anthropomorphic phantom, ADMIRE removed noise near edges more efficiently than SAFIRE.

**Conclusion:** No significant improvement in maintaining noise structure were found for the ADMIRE algorithm. Based on the inter-image standard deviation maps, ADMIRE removed noise near edges more efficiently than SAFIRE.

## 1 Purpose

Computed Tomography (CT) has for decades played an essential role in diagnostic radiology. Since introduced in the 1970s, the imaging modality has undergone great technological advances and the clinical applications have increased. Conventional CT image reconstruction is fast, gives consistent image quality yielding potential for clear and accurate clinical decision-making [1]. However, since X-rays in the relatively small doses given in diagnostic use might be connected to a small increase in cancer incidence, several dose-reduction strategies has been proposed in the last several years. The introduction of noise-reducing iterative reconstruction (IR) techniques is a recent and important contributor to the dose reduction in CT [2]. All vendors offer at least one IR solution on their CT systems. Sinogram Affirmed Iterative Reconstruction (SAFIRE) is an IR algorithm made available by Siemens Healthcare, and is available on most of their CT systems. Several studies have confirmed the dose reduction potential for SAFIRE [3-6]. Recently, their next-generation IR algorithm, called Advanced Modeled Iterative Reconstruction (ADMIRE), has become clinically available on high-end CT systems manufactured by Siemens Healthcare.

With the traditional filtered back projection (FBP) reconstruction algorithm, there is a well-known relationship between radiation dose, measured noise and perceived diagnostic image quality; Noise increases as the inverse square root of the dose. Most IR algorithms are nonlinear in nature, and through their application, low dose does not necessarily translate into high noise [7]. While the application of IR during tomographic reconstruction reduces image noise, it has also been shown that IR algorithms push the noise structure towards the lower end of the Fourier spectrum, resulting in a "blotchy" appearance of the noise in the presented images [8]. Due to the algorithmic complexity of identifying and reducing noise in high-contrast regions such as organ edges, the IR algorithms preserve more of the noise in such border regions compared to homogeneous areas to maintain spatial resolution. This results in a heterogeneous noise reduction for clinical images where IR is applied [4, 9].

The ADMIRE algorithm is classified as a Model Based Iterative Reconstruction (MBIR) algorithm, which utilizes more advanced noise reduction methods, such as system and noise modelling [2], compared to traditional IR algorithms. According to the vendor, the statistical modeling in the projection and image domains has been substantially improved for ADMIRE compared to SAFIRE. ADMIRE is supposed to improve the preservation of anatomic edges in the patient as well as keep the noise texture unchanged relative to FBP, due to the inclusion of a larger volume in the analysis of the neighboring voxel data in the image domain [10]. Keeping the noise at a low level and at the same time preserving low-contrast details is critical for the diagnostic quality e.g. in the abdomen, where organs may contain low-contrast lesions.

Some studies have compared noise properties in images reconstructed with IR, compared to using FBP [4, 9]. However, to the authors' best knowledge, no studies comparing noise properties for two different IR algorithms from the same vendor have been performed. The purpose of the present study is to compare the noise properties originating from the two iterative algorithms, SAFIRE and ADMIRE, using an abdomen protocol at two different dose levels to evaluate if any differences in noise structure and noise reduction properties near anatomic edges could be found. If significant differences between these two algorithms exist, it might be worth the cost of upgrading, thereby achieving a lower dose exposure for the patients.

## 2 Methods and Materials

### 2.1 Image acquisition

The image acquisitions were performed on a Siemens Somatom Definition Flash (Siemens Healthcare, Forchheim, Germany) with the ADMIRE algorithm and a Siemens Somatom Definition AS+ with the SAFIRE algorithm. The scans were performed on the image quality phantom Catphan 600 (The Phantom Laboratory, Salem, NY) with an elliptical annulus Catphan CTP579 attached to mimic the human abdomen in size and shape. The Catphan phantom includes several modules for image quality assessment. In this study the uniformity module CTP486 was utilized to perform noise measurements. In order to obtain anatomic images for the study, an anthropomorphic abdomen phantom PH-5 CT Abdomen Phantom (Kyoto Kagaku, Japan) was scanned: The two phantoms are shown in Figure 2.

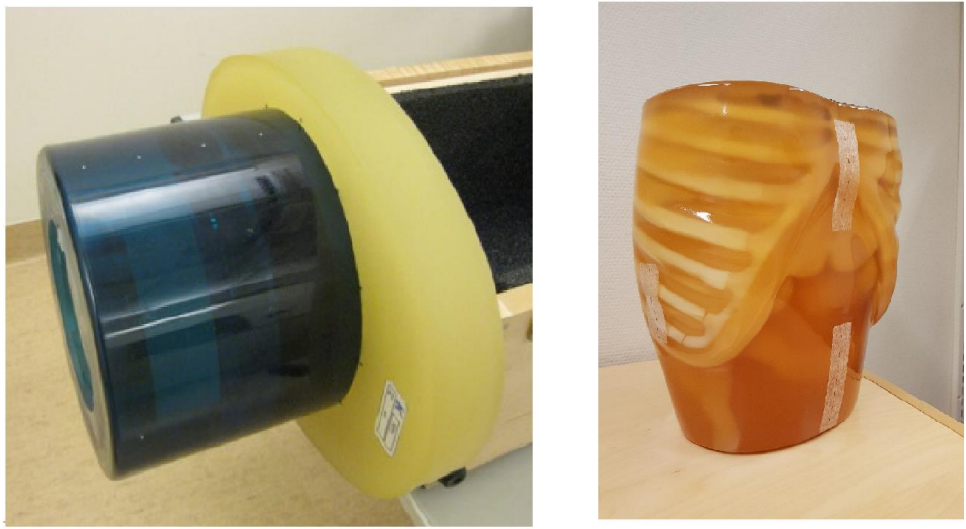

**Figure 1** Left: The Catphan 600 phantom with the elliptic annulus CTP579 attached. Right: The PH-5 CT Abdomen Phantom

All scans were performed in axial mode using a constant tube current and an abdomen protocol. These scan settings were chosen to simplify the analysis: The constant tube current ensures that the analysis of the resulting image noise is easier to perform. The axial mode was selected to ensure that the scan table is stationary, producing similar slice positions.

Two different dose levels were selected: a "full dose" level representing a typical $CTDI_{vol}$ for the abdomen area ($CTDI_{vol}$ = 10 mGy) and a "low dose" level at 20% of the full dose level ($CTDI_{vol}$ = 2 mGy). All scans were taken at 120 kV. A medium soft abdomen filter, B30, was used to reconstruct all acquired images. All scans were reconstructed with 3 different strengths of the IR algorithm; low (level 1), medium (level 3) and high (level 5), in addition to traditional FBP (defined as level 0): see Table 1 for overview. Both algorithms have five different strengths in total.

### 2.2 Noise measurements using Catphan

The *noise texture* is a term used to explain the appearance of noise in the reconstructed image slices. The texture may be defined in terms of the spectral components of the noise: In order to visualize the spectral distribution of the image noise, the Noise Power Spectrum (NPS) is calculated from a number of similar images taken at the same slice position.

Table 1: An overview of the acquisition and reconstruction parameters used in this study

| Scanner | Siemens Somatom Definition AS+ | | Siemens Somatom Definition Flash | |
|---|---|---|---|---|
| **Abdomen protocol** | "Full dose" | "Low dose" | "Full dose" | "Low dose" |
| **Tube current product (mAs)** | 150 | 30 | 150 | 30 |
| **$CTDI_{vol}$ (mGy)** | 10 | 2 | 10 | 2 |
| **Tube potential (kVp)** | 120 | 120 | 120 | 120 |
| **Kernel** | B30f | B30f | B30f | B30f |
| **Iterative strength** | 0 (FBP), 1, 3, 5 | 0 (FBP), 1, 3, 5 | 0 (FBP), 1, 3, 5 | 0 (FBP), 1, 3, 5 |
| **IR algorithm** | SAFIRE | SAFIRE | ADMIRE | ADMIRE |

Fifty images were acquired (five scans, ten noise images from each acquisition) of the Catphan phantom module CTP486 in order to calculate the NPS for all combinations of dose levels and IR levels on both CT scanners [11]. The number of images was chosen for a high statistical basis for the NPS calculations: the result being smoother spectral curves. A 2D polynomial was subtracted from all images in order to remove the residual low-frequent signals resulting from beam hardening (cupping artifact). The resulting NPS curves show the noise structure through its distribution in the frequency domain. All curves were normalized in area to easier visualize changes in the noise distribution. Finding a single metric is helpful for comparison between different noise curves, and to this end, the median values of the curves were calculated. Changes in noise structure due to the application of IR filters can be identified through changes in the median value of the NPS curve.

In addition to the median NPS value, another metric was calculated as a cross-validation for the method. To this end, the noise was measured in terms of standard deviations of the Hounsfield Units (HU) in a ROI: an average from 50 noise images was used.

## 2.3 Inter-image standard deviation using an anthropomorphic phantom

The NPS method is used to identify the spectral structure of the noise resulting from different scan settings. In addition, it is helpful to visualize the *spatial* distribution of noise in order to understand how ROIs within the reconstructed image slices are affected by the different scan settings: This enables a visual assessment of the results of the applied IR filters.

The anthropomorphic phantom was scanned 50 times at the exact same table position, yielding 50 CT slices varying only in the random fluctuations of the noise. Based on these 50 scans, the standard deviation in each pixel was calculated. This was performed by finding the sequence of HU values originating from the same pixel position in all acquired images, and calculating the standard deviation of these values: The result is a so-called "inter-image standard deviation" of each pixel. The strength of this method is that the image noise may be more accurately spatially defined compared to calculating the HU standard deviation in a ROI [12]. Two-dimensional maps of the inter-image standard deviation, i.e. for all the pixels in the image, were calculated for all combinations of dose levels and IR levels on both CT scanners.

# 3 Results

## 3.1 Noise Power Spectra and noise

For both ADMIRE and SAFIRE, the NPS curves are shifted towards lower spatial frequencies with increasing IR level (4% - 31%) at both dose levels.

The NPS curves and median values are shown for all reconstructed images in Figure 3. At the "full dose" acquisitions ($CTDI_{vol}$ = 10 mGy), the median values of the NPS curves were similar for both scanners at all IR levels (0, 1, 3 and 5). The noise magnitude (standard deviation) was similar for all IR levels.

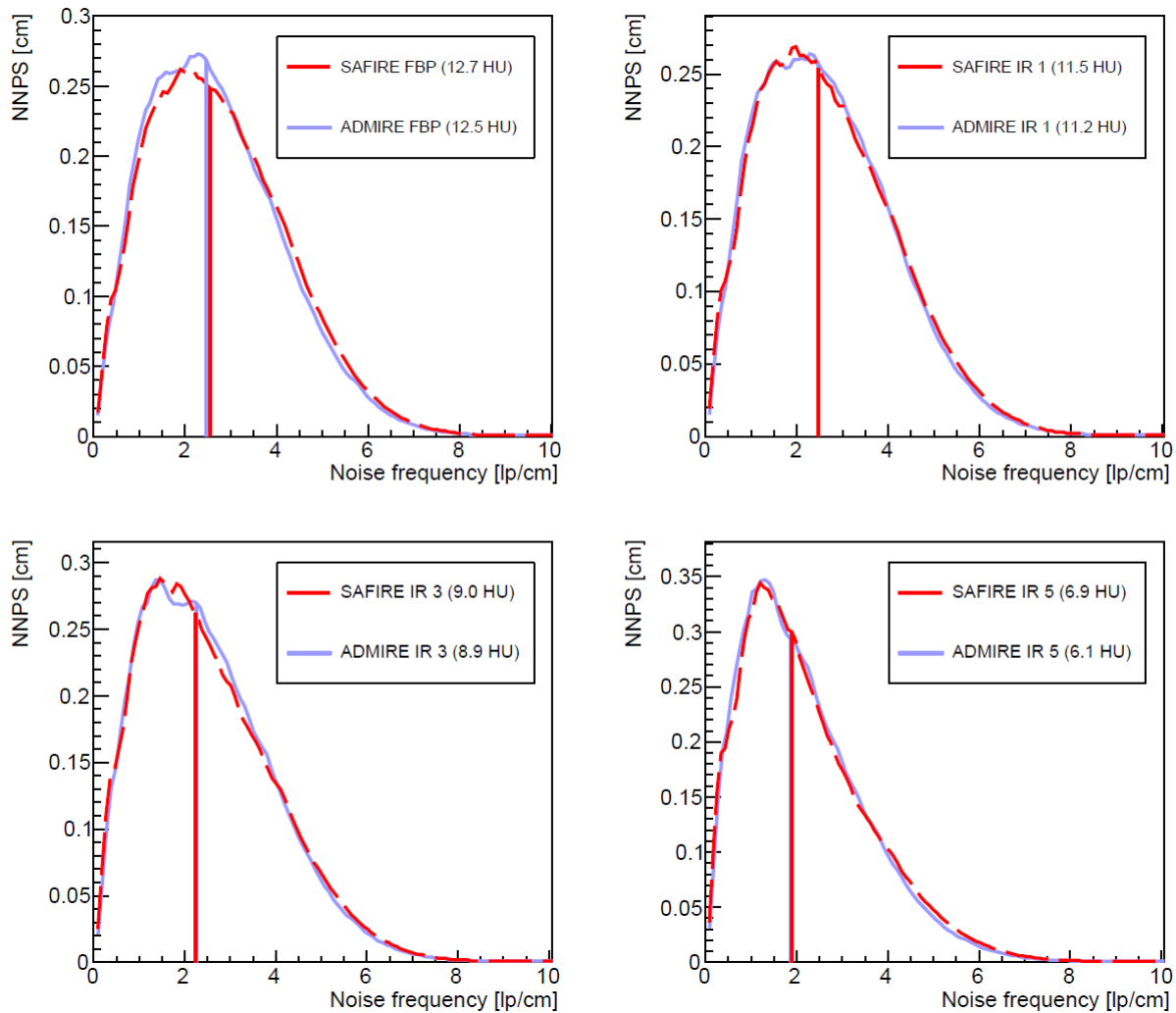

**Figure 2: The figure shows normalized Noise Power Spectrum curves for pure noise images at full dose (CTDI = 10 mGy) reconstructed with a medium sharp abdomen filter (B30) with FBP and iterative level 1,3 and 5 of SAFIRE using Siemens Somatom Definition AS+ (red curves) and ADMIRE using Siemens Somatom Flash (blue curves). The noise in terms of the HU standard deviation is given for each curve in the legend.**

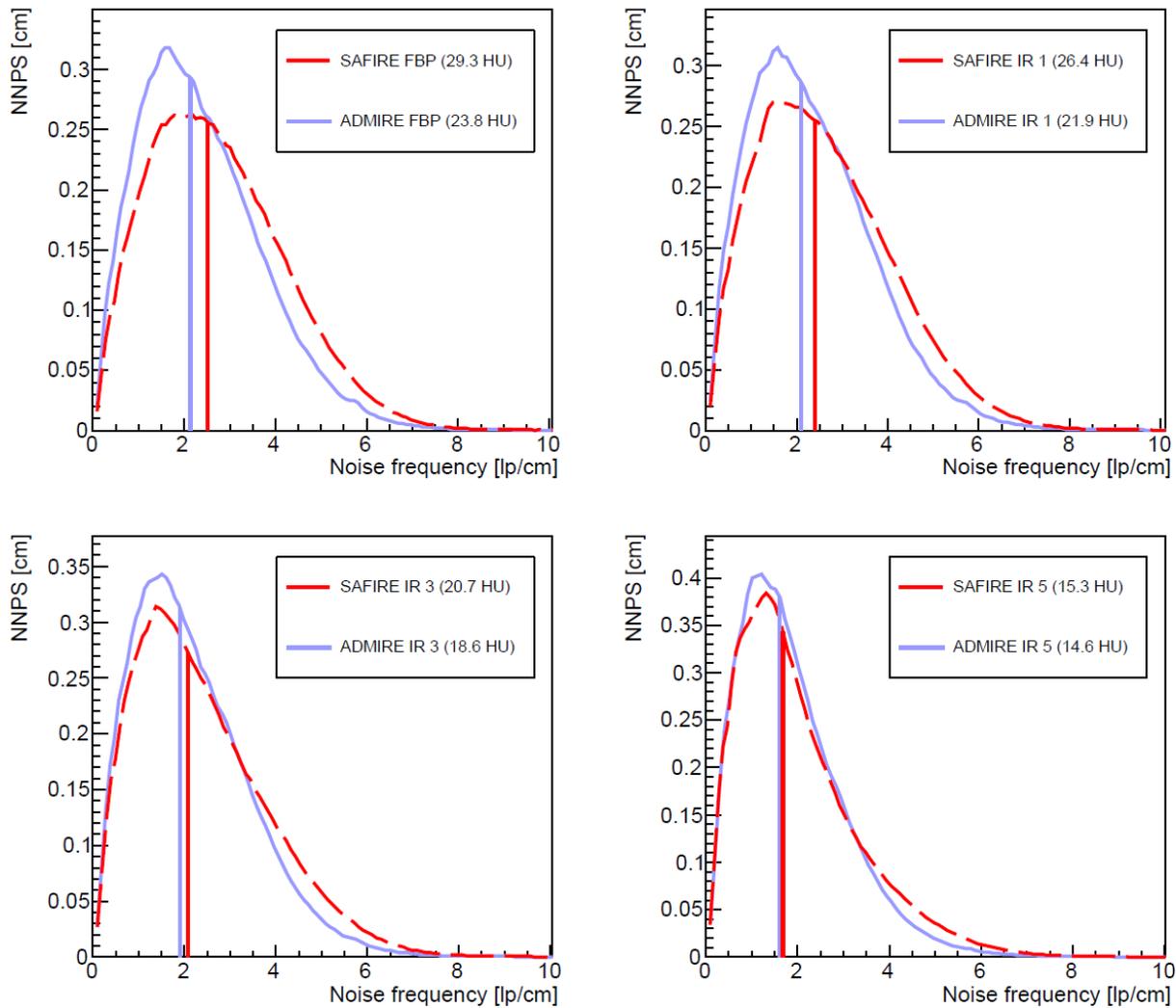

**Figure 3:** The figure shows normalized Noise Power Spectrum curves for pure noise images at low dose reconstructed with a medium sharp abdomen filter (B30) with FBP and iterative level 3 and 5 of SAFIRE using Siemens Somatom Definition AS+ (red curves) and ADMIRE using Siemens Somatom Flash (blue curves). The noise in terms of the HU standard deviation is given for each curve in the legend.

For the scans acquired at low dose ($CTDI_{vol}$ = 2 mGy), the median values of the NPS curves are generally shifted towards lower spatial frequencies. This shift is more prominent for ADMIRE compared to SAFIRE for all IR levels (0, 1, 3 and 5). The NPS median values are 14%, 12 %, 9% and 4% less for ADMIRE than SAFIRE, respectively, for IR levels 0, 1, 3 and 5. Comparing the noise magnitudes, the analysis shows that ADMIRE noise levels are below the measured noise in SAFIRE: For ADMIRE the measured noise is 19%, 17%, 10% and 5% less than for the AS+ scanner with SAFIRE, respectively, for IR levels 0, 1, 3 and 5.

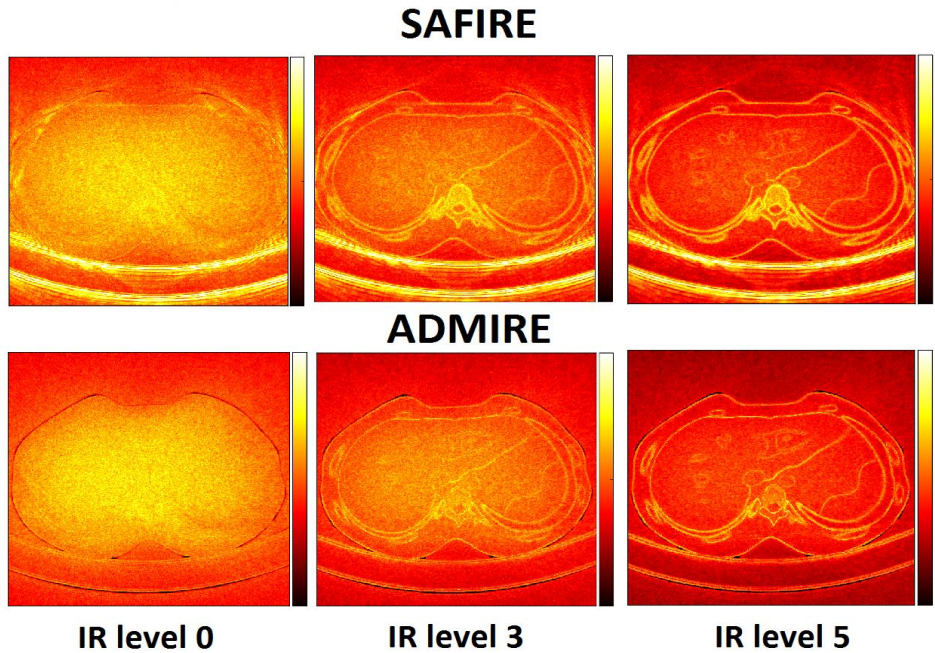

**Figure 4:** Inter-image standard deviation maps show how noise (in magnitude) is distributed throughout the anthropomorphic phantom for FBP (IR level 0) and iterative level 3 and 5 at identical slice location in the phantom at both scanners. The upper images are from the Definition AS+ scanner using SAFIRE, while the lower images are from the Definition Flash scanner reconstructed with ADMIRE. These images are taken at full dose. Increasing brightness indicate higher noise magnitudes.

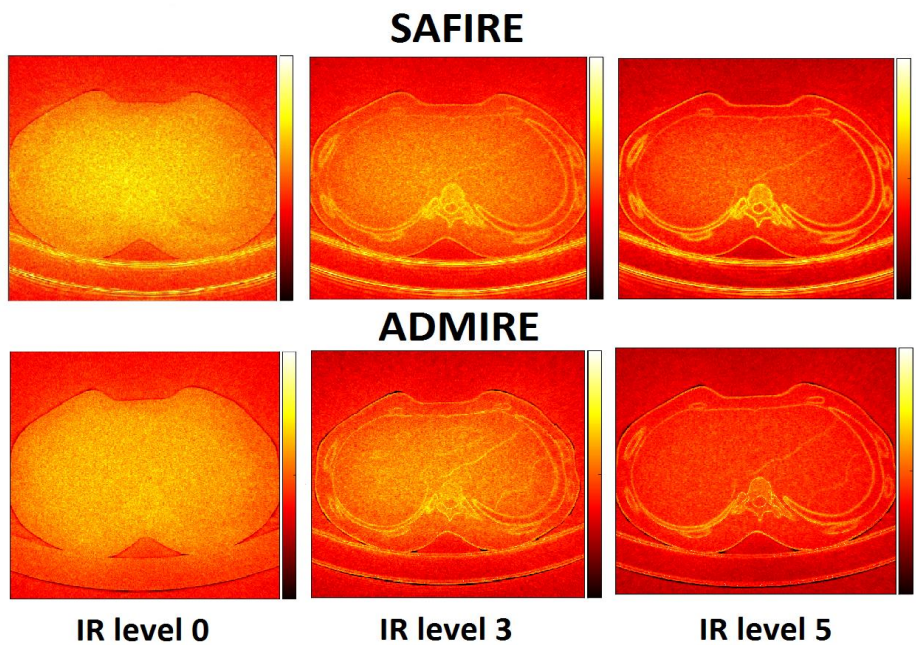

**Figure 5:** Inter-image standard deviation maps show how noise (in magnitude) is distributed throughout the anthropomorphic phantom for FBP (IR level 0) and iterative level 3 and 5 at low dose ($CTDI_{vol}$ = 2 mGy). The upper images are from the Definition AS+ scanner using SAFIRE, while the lower images are from the Definition Flash scanner reconstructed with ADMIRE.

## 3.2 Inter-standard deviation maps

The inter-standard deviation maps are shown in Figure 4 for full dose (CTDI$_{vol}$ = 10 mGy) and in Figure 5 for low dose (CTDI$_{vol}$ = 2 mGy). For the FBP images, the noise is evenly distributed within the phantom. For the IR images, the noise is heterogeneous, especially near high contrast edges like bone and near edges of low contrast. For both IR algorithms the noise in areas close to edges is higher than the noise in homogenous regions. However, ADMIRE reduces the noise closer to all edges than SAFIRE. This can especially be seen in the spine where the noise near edges are less blurry in images reconstructed with ADMIRE. Noise in homogenous regions in the phantom was significantly reduced with increasing IR strength for both IR algorithms. As can be seen in the spatial distribution of the noise in Figures 4 and 5, the inter-image standard deviation maps show greater noise reduction in homogenous areas with ADMIRE at low dose as compared to SAFIRE. In the noise maps reconstructed with FBP (IR level 0), there is less noise for all scans taken at the Somatom Flash scanner with the ADMIRE algorithm.

## 4 Discussion

We have found both similarities and differences in the noise properties of the SAFIRE compared to the ADMIRE IR algorithms. The general shift in NPS curves has been shown in previous studies [8, 13, 14], where it has been shown that IR algorithms not only reduce noise, but also affects the noise structure as evidenced by a shift towards lower noise frequencies in the Noise Power spectra.

A significant difference in measured noise magnitude (STD) were found, with ADMIRE having lower values for all IR levels (0, 1, 3 and 5). For the images acquired using "full dose", no significant differences in median NPS values were found at the various IR levels, indicating similar noise structure properties for the two different IR algorithms. For the images acquired using "low dose", both IR systems were seen to shift the NPS curves towards the lower frequencies: The shift distance was greater in the ADMIRE dataset compared to the SAFIRE dataset. This indicates that both of the studied IR algorithms are applied more aggressively when the overall level of noise is higher, and less aggressively when there is little noise.

The changes in the NPS curves and noise magnitudes were most prominent using IR level 0 (FBP). According to the vendor, even when applying only the FBP reconstruction, some statistical improvements are still performed. The shift in NPS values towards lower spatial frequencies and the overall reduced noise level for the Flash scanner with the ADMIRE algorithm may indicate that there is a stronger noise reduction and more statistical improvements for this algorithm. The noise reduction may also be partly explained by the fact that the two scanners are equipped with different detector systems, with the Flash scanner having the STELLAR detector which generates less electronic noise.

Both algorithms showed shifts in median NPS values in the order of up to 23 – 31% for the highest IR level compared to FBP (IR level 0). This result is somewhat different than found in a previous study on ADMIRE [9] where NPS peak values were studied at different dose levels and mostly showed a shift in the range of 0.5 – 2%. However, NPS peak values are not sensitive to the shape of the NPS curve, as the median values are, which may be part of the explanation of the different results. In addition this previously study was performed at a Siemens Somatom Force scanner, which has been installed with new kernels that are not available at the Flash scanner.

Based on the inter-image standard deviation maps calculated for the images of the anthropomorphic phantom, ADMIRE reduced noise near edges more efficiently than SAFIRE which may be due to the farther reaching neighborhood analysis of voxel data in the image domain that is performed by ADMIRE. The spatial distribution of noise with increasing IR level is consistent with what has been shown in a previous study [4] on noise properties for CT images with structures. In our study an abdomen phantom with mostly low contrast details and organs has been used, the results could be different, or even more prominent when using for instance a lung phantom with a sharper kernel.

It should be mentioned that even anthropomorphic phantoms like the one used in this study is fairly homogenous within each organ, lacking anatomic texture and detailed anatomical features which in real life could influence the performance of the IR algorithm and the final image quality [4]. To verify any significant differences in image quality performance between these two IR algorithms, real patient images should be used.

# 5   Conclusion

We have demonstrated both differences and similarities between the noise properties of images reconstructed by the SAFIRE and ADMIRE IR algorithms. No significant improvement in maintaining noise structure (i.e. shifts in the NPS curves) were found for the ADMIRE algorithm. However, the ADMIRE algorithm removes significantly more noise at low dose compared to SAFIRE, most prominently when using FBP: This difference might be explained by the different detector systems available for the two CT scanners.

The heterogeneous distribution of noise with increasing IR level is consistent with what has been shown in previous studies [4, 9] on noise properties for CT images with structures. Based on the inter-image standard deviation maps calculated for the images of the anthropomorphic phantom, ADMIRE removed noise near edges more efficiently than SAFIRE.

In our study we have shown that ADMIRE do not preserve noise structure better than SAFIRE, however the new IR algorithm reduces noise near edges better than SAFIRE.